\documentclass[3p,times,procedia]{elsarticle}
\usepackage{nupha_ecrc}

\volume{00}

\firstpage{1}

\journalname{Nuclear Physics A}

\runauth{T. Sj\"ostrand}

\jid{nupha}

\jnltitlelogo{Nuclear Physics A}

\usepackage{amssymb}

\usepackage[figuresright]{rotating}

\begin{document}

\begin{frontmatter}



\dochead{\hfill LU TP 18-24\\ \hfill MCnet-18-18 \\ \hfill August 2018\\[3mm]
XXVIIth International Conference on Ultrarelativistic Nucleus-Nucleus Collisions\\ (Quark Matter 2018)}

\title{Collective Effects: the viewpoint of HEP MC codes}


\author{Torbj\"orn Sj\"ostrand}

\address{Department of Astronomy and Theoretical Physics, Lund University,
S\"olvegatan 14A, SE-223 62 Lund, Sweden}

\begin{abstract}
Collective effects are observed in high-multiplicity pp events, 
similar to the signals traditionally attributed to the formation 
of a Quark Gluon Plasma in heavy ion collisions. In core--corona models 
it is assumed that a partial plasma formation is indeed possible also in
pp, but here the focus is on several recent models that attempt to  
explain pp data without invoking plasma formation. These attempts are
partly successful, but there is still not a unified framework.
\end{abstract}


\end{frontmatter}


\section{Introduction}

The relationship between the pp and the AA communities at the LHC
is changing. This has been brought about by a set of unexpected 
observations, wherein high-multiplicity pp events seem to attach 
smoothly to the behaviour observed in pA and AA collisions, with 
respect to flavour composition \cite{ALICE:2017jyt} and flow
\cite{Khachatryan:2010gv,Khachatryan:2015lva, Aad:2015gqa,%
Ortiz:2013yxa,Abelev:2014qqa,Khachatryan:2016txc}.  

For pp we have been used to a simple picture of hadronization,
wherein the density of colour fields and hadrons has been assumed
low enough that interactions between them can be neglected. Then
e$^+$e$^-$ and pp events should share many common traits,
``jet universality''. This picture is embedded in the greater
context of event generators \cite{Buckley:2011ms}, wherein also all 
other aspects of pp events are simulated. These programs provide an 
overall description of event properties that in the past appeared 
reasonably successful, and that still today can describe the bulk of 
data distributions quite well.

For the AA community the key concept has instead been the Quark Gluon 
Plasma (QGP), how it is created, what properties it has, and how it
reverts back to ordinary matter. Catchwords include deconfinement,
hydrodynamics, flow, perfect liquid, and more, all unknown in 
traditional pp approaches. This split has been encouraged by standard
QGP theory, where the belief has been that pp collisions cannot 
generate a sufficiently large volume sufficiently long for a QGP to form
\cite{BraunMunzinger:2007zz,Busza:2018rrf,Nagle:2018nvi}.

Thus both communities have acted to keep a barrier between pp and AA
physics. But now it is time to open up the discussion and ask some
tough questions. Is a QGP formed in high-multiplicity pp events?
If not, what other mechanisms could one imagine as being at the origin
of the observed pp behaviour? How can these be tested? Specifically,
which are the ironclad signals of QGP formation? 
  
Answering these questions will keep us busy in the years to come.
In this talk some of the attempts already made will be discussed,
within the context of the traditional event generators used to 
describe pp collisions. Other presentations at this conference provide 
the view from other vantage points.

\section{pp physics and generators}

Hadronization is traditionally assumed to be environment-independent. 
Since it is a nonperturbative process, free parameters are needed for 
incalculable quantities, but these can be determined e.g.\ from LEP 
e$^+$e$^-$ data and then be applied unchanged for LHC pp collisions. 
The modelling of the partonic state that is to hadronize has to be 
different in the two processes, of course. In e$^+$e$^-$ events only
final-state radiation and hadronization needs to be considered,
while the composite nature of the proton additionally leads to parton 
distribution functions, initial-state radiation, beam remnants, and 
multiparton interactions (MPIs) \cite{Sjostrand:2017cdm}. 

MPIs imply several subcollisions in an average pp event, typically 
with the outgoing scattered partons having $p_{\perp}$ scales of a few 
GeV and being colour-connected to the beam remnants. Thereby a number 
of colour confinement fields --- strings \cite{Andersson:1983ia} --- 
are stretched essentially longitudinally between the two remnants. 
A rule of thumb is that a single string gives about one charged 
particle per unit of rapidity, i.e.\ a typical LHC value 
$\langle \mathrm{d}n_{\mathrm{charged}} / \mathrm{d}y \rangle \approx 6$ 
would correspond to the order of six strings being pulled out. 
It is also useful to note that the tail to higher multiplicities is
driven by events with many MPIs, rather than e.g.\ events with a 
pair of high-$p_{\perp}$ jets.

In such a picture there are no collective effects of any importance. 
The observation of a rising $\langle p_{\perp} \rangle(n_{\mathrm{charged}})$ 
at the Sp$\overline{\mathrm{p}}$S was addressed by the introduction of
Colour Reconnection (CR), however \cite{Sjostrand:1987su}. Here the 
colour fields of the event can be redirected, relative to the naive 
picture of colour-separated MPIs, in such a way that the total string 
length is reduced. Several different CR scenarios have been proposed, 
and for each such the $\mathcal{P}(n_{\mathrm{charged}})$,
$\langle p_{\perp} \rangle(n_{\mathrm{charged}})$ and other data can
be used to tune free parameters. CR can give some effects that are of
a collective-flow character, by providing transverse boosts to 
reconnected string pieces, that e.g.\ gives heavier hadrons higher 
$\langle p_{\perp} \rangle$, but CR does not address many other issues. 

Currently the most successful realistic approach to collective effects 
in pp is the core--corona one, as implemented in the EPOS event generator 
\cite{Pierog:2013ria,Werner:2013tya}. In it the MPIs give rise to 
(mainly) longitudinally stretched strings. As long as these strings 
are well separated they hadronize independently, a ``corona'',  but in 
case of close-packing they are assumed to collectively give rise 
to a local QGP, a ``core'', which expands according to hydrodynamics 
and hadronizes according to a statistical model. An event can be a 
mixture of the two. In low-multiplicity events only the corona may 
exist, but with increasing multiplicity the core fraction increases. 
EPOS works not only for pp events, but extends the same formalism to 
pA and AA, where the core QGP component dominates.   
Obviously this formalism is very economical, in that it  does not 
require the introduction of any new principles. A smooth transition
between two extreme behaviours is obtained by changing the admixture,
but the core and corona components in a given event are discontinuously 
separated. EPOS is primarily a model for soft physics, however, not
for hard processes and the underlying events associated with them, 
and therefore is not suited for much of the pp studies at the LHC. 

The physics of EPOS is already well known within the AA community,
and therefore it will not be discussed any further here. Instead the
attention will be turned to other models that have been proposed, 
specifically within the context of the event generators normally used
for pp studies. These are especially interesting insofar as they 
do not assume the formation of a QGP, but instead introduce alternative
physics mechanisms that could be at play. The flip side is that these
models are not yet extended to pA and AA collisions. 

The three main pp event generators are 
\textsc{Pythia} \cite{Sjostrand:2006za,Sjostrand:2014zea}, 
Herwig \cite{Bahr:2008pv,Bellm:2015jjp} 
and \textsc{Sherpa} \cite{Gleisberg:2008ta}. 
While sharing the same overall common structure,
there are still many physics differences, and philosophy ones.
\begin{itemize}
\item \textsc{Pythia} has its roots back in the string fragmentation
studies in the late 70ies. A string can stretch e.g.\ from a quark end 
via a number of (colour-ordered) intermediate gluons to an antiquark 
end, and fragments along its full length \cite{Sjostrand:1984ic}.
Studies of soft physics have always been central, like MPIs and CR 
\cite{Sjostrand:2017cdm}. The Fritiof model for AA collisions 
\cite{Andersson:1986gw,NilssonAlmqvist:1986rx} was a separate offshoot, 
but recently the  related Angantyr pA/AA model is a fully integrated 
part of \textsc{Pythia} \cite{Bierlich:2018xfw}. Many other event 
generators are built on top of \textsc{Pythia} code, and so are most 
of the alternative scenarios to be discussed here.
\item Herwig was begun in the mid-80ies, to study coherent parton shower
evolution \cite{Marchesini:1983bm}. Hadronization is based on cluster 
fragmentation \cite{Webber:1983if}, wherein gluons are forced to break
up into quark--antiquark pairs at the end of the shower, such that 
lower-mass clusters replace the long strings. Both MPIs and CR are
modelled, but along somewhat different lines than in \textsc{Pythia}.
\item \textsc{Sherpa} grew out of matrix-element generator activities
in the late 90ies, and the focus of attention has been in the matching 
and merging of matrix elements and parton showers. Cluster fragmentation 
is default, but the program can be linked to \textsc{Pythia} for string 
fragmentation. The default MPI/CR machinery is inspired by the 
\textsc{Pythia} one, but the KMR model implemented in the 
SHRiMPS code will be made available as an alternative \cite{Martin:2012nm}.
\end{itemize} 
 
\section{Flavour composition}

A significant strangeness enhancement is observed in high-multiplicity
pp events \cite{ALICE:2017jyt}. This is visible in K$^0_{\mathrm{S}}$ and
$\Lambda$ production, but in particular in the multistrange $\Xi$ and
$\Omega$ production. The proton fraction, on the other hand, remains
fairly constant, so the effect does not appear to be related to baryon
number or particle mass. ALICE shows that this phenomenon is not described 
by the standard string fragmentation framework, which has an essentially 
multiplicity-independent particle composition, while the core--corona
model in EPOS has the right trends but overshoots, and the rope model in
DIPSY/\textsc{Pythia} \cite{Bierlich:2014xba} provides a decent description.
Let us study expectations further. 

In the standard string model, the string tension $\kappa$ is assumed to 
be a constant, $\kappa \approx 1$~GeV/fm. When such a string is pulled 
out between two receding colour charges, it can break by the production 
of a quark--antiquark pair that screens the endpoint colour charges.    
Such a break can be viewed as a tunneling process, where the pair is 
created in one common point but then q and $\overline{\mathrm{q}}$ each 
has to tunnel out a distance $d = m_{\perp\mathrm{q}} / \kappa$ to become 
on shell, where $m_{\perp\mathrm{q}} = m_{\perp\overline{\mathrm{q}}}$ is the 
transverse mass of the quark. This gives a relative probability
\begin{equation}
\mathcal{P} \propto 
\exp \left( - \frac{ \pi m_{\perp\mathrm{q}}^2}{\kappa} \right) 
= \exp \left( - \frac{ \pi p_{\perp\mathrm{q}}^2}{\kappa} \right) 
\times \exp \left( - \frac{ \pi m_{\mathrm{q}}^2}{\kappa} \right) ~,
\label{eq:tunnel}  
\end{equation}
i.e.\ a common Gaussian $p_{\perp}$ spectrum for all quarks, 
and a suppression of the production of heavy quarks. Quark masses 
are ill-defined, so the strangeness suppression is viewed as a free 
parameter, of the order of 0.2 - 0.25, while charm and bottom are 
so suppressed that their nonperturbative production can be neglected.
Other aspects also influence the meson production, such as the 
relative rate of pseudoscalars and vectors.

The real problem is baryon production. In the simplest approach 
a colour antitriplet diquark is viewed as equivalent to an antiquark,
and produced by the same tunneling process as above. Different diquarks
are again suppressed in relation to their squared masses, with some free
parameters to represent the uncertainty in diquark mass patterns.
Unfortunately the diquark model gives too strong a suppression of 
multistrange and spin-$3/2$ baryons. An extension is the popcorn model
\cite{Andersson:1984af}, wherein quark--antiquark pairs are created 
one at a time, and the suppression of rare hadrons is not as extreme, but 
still too large e.g.\ for $\Omega$ production. There are also problems 
e.g.\ with azimuthal correlations in baryon pairs \cite{Adam:2016iwf}, 
so it is clear we still lack some fundamental insight on baryon 
production, at least in the string context.

The transverse size of a string is of the order of the proton radius.
Therefore, when two protons collide and several strings are formed 
by MPIs, it is almost unavoidable that these strings come to overlap
in space--time. This has been used as an argument for colour reconnection, 
but otherwise the possibility of collective effects has largely been 
neglected. In the past few years some explicit models have appeared, 
however. 
\begin{itemize}
\item The rope model \cite{Biro:1984cf,Bialas:1984ye,Bierlich:2014xba} 
assumes that several nearby strings can be intertwined into a rope, 
which represents the field drawn out by the combination of several colour 
charges. Consider the example of two parallel strings, for which 
$3 \otimes 3 = 6 \oplus \overline{3}$, where the sextet has a  Casimir 
colour factor $C_2^{(6)} = \frac{5}{2}C_2^{(3)}$. In the first break of 
such a rope the effective string tension is proportional to 
$C_2^{(6)} - C_2^{(3)}$, i.e. $\kappa_{\mathrm{eff}} = \frac{3}{2} \kappa$
should be used in eq.~(\ref{eq:tunnel}).
For a second break (in the same region) the string tension is back to
the normal one. For multiple (almost) collinear strings one could 
expect some kind of random walk in colour space, allowing higher
colour charges to be reached, and thereby also larger 
$\kappa_{\mathrm{eff}}$. In those string breaks the mass suppression
of strangeness and baryon production would be reduced. The rope model 
describes the production of (multi)strange baryons fairly well, 
as already mentioned, but does predict a rise of the p$/\pi$ ratio
with increasing multiplicity, in conflict with data. 
\item Most simple CR models do not change the hadron composition, but a
QCD-colour-factor-based CR model does \cite{Christiansen:2015yqa}. 
Again consider the relation $3 \otimes 3 = 6 \oplus \overline{3}$, 
but now for the $\overline{3}$ possibility that 
two parallel strings may fuse to produce a normal string, although
with the colour flow in the opposite direction. Near either endpoint, 
where either two q or two $\overline{\mathrm{q}}$ are located, the 
fused string needs to split into two that stretch to the two endpoint
quarks, and the point of splitting is a so-called junction. 
It becomes associated with a baryon number, and an antijunction at the 
other end with an antibaryon. Since the number of reconnections 
increases faster than the number of individual strings, 
it means that the baryon fraction increases with multiplicity.  
Furthermore, since a junction baryon consists of the flavours 
produced at the three separate string breaks closest to the junction
in each of the three string legs out from it, production of multistrange 
baryons is not suppressed by a large strange diquark mass in the 
tunneling expression. Qualitatively it therefore describes the ALICE
trends of $\Xi$ and $\Omega$ being more common at high multiplicities, 
but unfortunately some of the rise is also present for p and $\Lambda$.
\item From ISR days (pp collisions up to $\sqrt{s} = 62$ GeV) it has 
been known that hadron production $p_{\perp}$ can be given a 
thermodynamical-like description, e.g.\ in terms of a mass-dependent
$p_{\perp}$ spectrum
\begin{equation}
\frac{\mathrm{d}\sigma}{\mathrm{d}^2 p_{\perp}} 
= N \, \exp \left( - \frac{m_{\perp \mathrm{had}}}{T} \right) ~~,~~
m_{\perp \mathrm{had}} = \sqrt{ m_{\mathrm{had}}^2 + p_{\perp}^2} ~,
\end{equation}
where $N$ and $T$ are (approximately) common for all hadrons. 
An effectively exponential fall-off could arise also starting from the
Gaussian one in eq.~(\ref{eq:tunnel}), assuming that the string tension 
is fluctuating along its length, also in the absence of other strings
\cite{Bialas:1999zg}. An option has been added to the \textsc{Pythia}
string model based on an exponential suppression, but with local flavour 
and $p_{\perp}$ conservation \cite{Fischer:2016zzs}. Such an ansatz
gives an overall decent description of the particle composition with only
a few free parameters, but does overestimate the rate of multistrange 
baryons. A variable string tension or ``temperature'' is used in cases 
of close-packing of strings, with a continuous change as strings become 
squeezed into smaller transverse areas, with results similar to those 
of the (discrete-step) rope model.
\end{itemize}
In summary, a few different ways have been introduced whereby the string 
model can be made to display a rising trend of multistrange baryon 
production. All of them share the problem that this rise inherently is 
accompanied by a rise of the overall baryon production rate, in
contradiction with ALICE data. 
  
In Herwig the cluster model has been improved in two ways 
\cite{Gieseke:2017clv}. Firstly, if three quark--antiquark clusters 
are aligned in parallel, then the three quarks can reconnect to a baryon 
cluster, and the three antiquarks to an antibaryon one. Secondly, 
nonperturbative g$\to$s$\overline{\mathrm{s}}$ branchings have 
been introduced, in addition to the conventional
g$\to$u$\overline{\mathrm{u}}$ and g$\to$d$\overline{\mathrm{d}}$ ones.
Together these two changes gives significant improvements in a 
number of respects, such as the K/$\pi$ and p/$\pi$ $p_{\perp}$ spectra.  
The rate of $\Lambda$, $\Xi$ and $\Omega$ production is significantly 
increased, even if still below data. The fraction of strange baryons 
increases with multiplicity, since the chance of baryon reconnections
increases in events with many MPIs, but unfortunately then so does 
the fraction of protons, just like for the \textsc{Pythia} modifications.

\section{Collectivity and flow}

Colour Reconnection can induce some of the signals often attributed to 
collectivity. The rising $\langle p_{\perp} \rangle(n_{\mathrm{charged}})$ 
is a prime example. Without the CR, the event multiplicity would rise
approximately proportionally to the number of MPIs, but with CR each
further MPI contributes successively fewer further hadrons. Hence
the perturbative $p_{\perp}$ associated with the MPIs is be shared 
among fewer hadrons, giving a larger average.

Heavier hadrons also have harder $p_{\perp}$ spectra, with 
K/$\pi$ and p/$\pi$ $p_{\perp}$ yield ratios that increase rapidly from 
almost zero at low $p_{\perp}$ scales to maximal at around 2--3~GeV
\cite{Sirunyan:2017zmn,Adam:2015qaa}, in decent agreement with EPOS and 
\textsc{Pythia}. Three effects contribute: resonance decays, which tends 
to produce many $\pi$ at small $p_{\perp}$ values, (mini)jet fragmentation, 
wherein heavier hadrons take a larger fraction of the parton momentum, 
and transverse string (or cluster) boosts, wherein a string piece moving 
with a fix transverse velocity will impart that velocity (on the average) 
to the hadrons produced from its fragmentation. The latter mechanism  
is enhanced significantly by CR \cite{Ortiz:2013yxa}. The p/$\pi$ 
fraction drops above the peak position, and this drop is underestimated
e.g.\ in \textsc{Pythia} \cite{Ortiz:2013yxa}, suggesting that the 
transition to baryon production in jets is not so well modelled. 

In a related characterization, the $\langle p_{\perp} \rangle$ is rising
significantly as a function of the hadron mass \cite{Abelev:2014qqa},
which could be interpreted as a sign of collective flow. The same three
mechanisms as above combine to produce a decent description in 
\textsc{Pythia} and \textsc{Sherpa}. The trend is somewhat 
underestimated, however, and this gap is difficult to close
\cite{Fischer:2016zzs}. A model with an exponential $m_{\perp}$ spectrum,
e.g., intrinsically does give a steeper 
$\langle p_{\perp} \rangle (m_{\mathrm{had}})$ than the Gaussian $p_{\perp}$
default, and so should fit better. But most pions come from decays of 
heavier particles and thus pion changes are more related to the $m_{\rho}$ 
scale than to the $m_{\pi}$ one, thereby suppressing differences. Likely 
some reasonable amount of hadronic 
rescattering in the final state is needed to bring agreement between data 
and models. This is already standard in AA generators, using programs 
such as UrQMD \cite{Bass:1998ca} or SMASH \cite{Weil:2016zrk}. As a first 
step, the space--time production process in string fragmentation has 
recently been mapped out \cite{Ferreres:2018}, confirming that indeed 
hadrons are produced very closely packed.
   
One of the most spectacular signals of collective behaviour is the 
ridge effect \cite{Khachatryan:2010gv,Khachatryan:2015lva, Aad:2015gqa},
which is not predicted in conventional pp models. It is possible to obtain
a realistic model \cite{Bierlich:2016vgw} based on the concept of 
shoving \cite{Abramovsky:1988zh}, i.e.\ that strings that overlap in 
space--time repel each other and thereby build up a transverse velocity.
Initially strings are assumed to have zero width, but as they are pulled 
out in the longitudinal direction they also grow towards full transverse 
size. Therefore shoving always start at the middle of the local 
longitudinal rest frame and spreads outwards. For a practical 
implementation the event is sliced up into one unit wide rapidity 
ranges, and the net amount of shove on each string spanning that range 
is calculated. The resulting transverse momentum kick --- balanced so as 
to preserve total momentum --- is represented by a single gluon located
along the string at the relevant rapidity. The shoving effects on azimuthal
correlations become more important at higher multiplicities, in good 
agreement with data.   

Collective flow can be characterized by the commonly-used $v_n$ coefficient, 
which are non-vanishing not only in AA but also in pp collisions
\cite{Aad:2015gqa,Khachatryan:2016txc}. Part of this ``flow'' in pp 
comes from trivial sources, such as back-to-back (mini)jet pairs, but 
also colour reconnection and string shoving can contribute to the 
overall level of the signal e.g.\ in $v_2\{2\}$ \cite{Bierlich:2018lbp}. 
Unfortunately this signal is reduced appreciably when a gap is added 
to the $v_2\{2\}$ extraction, and the enhancement from CR is almost gone, 
while shove still gives a small positive contribution. Further studies 
are under way.

In  summary, we have bits and pieces of an understanding how collective 
flow can arise in pp also without a QGP, but not yet a complete view.

\section{Summary and outlook}

While a number of interesting and revealing studies have appeared in 
recent years, the multiplicity dependence of different pp event properties
could be investigated further. Are there signs of jet quenching at high 
multiplicities? A pattern of gradual 
$\Upsilon(1\mathrm{s}, 2\mathrm{s}, 3\mathrm{s})$ suppression?
A changing temperature of a soft prompt photon spectrum?
Changing flavour correlations, e.g.\ between baryons and antibaryons?
Is the flavour composition in jets more similar to e$^+$e$^-$ events 
or to the underlying event? And so on.

Some of these issues can be studied in the context of existing 
models coming from the AA side, such as core--corona ones. 
A shortcoming of implementations such as the EPOS generator is that 
these are focused on soft physics, and therefore do not currently 
have the full capability to  study the interplay between hard and soft 
aspects. This is where traditional pp event generators such as 
\textsc{Pythia}, Herwig and \textsc{Sherpa} may offer an advantage.
But for too long it was assumed that the combination of multiparton 
interactions and colour reconnection would generate all the collective 
effects needed to offer a reasonable description in pp, 
although we always knew that strings or clusters by necessity would 
come to be closely packed during the hadronization process, and that 
this should have repercussions. This is what we now try to remedy.
Effects can occur before the strings/clusters start to hadronize, 
exemplified by shoving or junction formation, during it, like ropes or 
a gradual change of the string tension, or after it, like hadronic 
rescattering. So far the formation of a QGP has not been invoked, 
however, so comparisons with EPOS will remain relevant where possible.

Quite apart from the close-packing issues, but probably exacerbated by 
them, current models for baryon production fail to provide a convincing 
description. The string model with a Gaussian suppression of quark and 
diquark masses, or for that matter an alternative with a Gaussian
suppression of hadron masses, suppresses the heavier multistrange 
baryons too much, while an alternative exponential formulation provides 
too little suppression. Correlations also are poorly described.

While we do have problems, recent and ongoing studies show that all is 
not hopeless. Some of the ideas do seem to provide a better understanding 
of data, but more is needed. It is also necessary to combine all the new 
pieces into a consistent framework. This could be achieved by upgrading
existing AA-style generators to provide more complete descriptions of
all kinds of pp event, or by extending pp generators to also simulate 
AA collisions. A simple stacking of (soft and hard) pp events here offers 
a possible starting point \cite{Bellm:2018sjt}, but the Angantyr 
model is somewhat more sophisticated, with further developments intended
to provide realistic descriptions of all pp/pA/AA collision types.

A key objective for current and future studies should be to better 
understand which experimental features are ironclad signatures of 
the formation of a quark gluon plasma, and which could be explained 
by other effects. The alternative explanations would likely also be of 
a collective character, like the models presented here, but not require 
a phase transition to another state of matter.

In summary, a whole new field of study has opened up in the last 
few years, and (seemingly?) made the borders between pp, pA and AA 
events crumble. Further experimental input will be crucial to understand 
what is going on, with model building in the context of event generators 
offering the main route to providing a global view of all possible effects.

\section*{Acknowledgement}

Work supported in part by the Swedish Research Council, contracts number
621-2013-4287 and 2016-05996, in part by the MCnetITN3 H2020 Marie 
Curie Initial Training Network, grant agreement 722104, and in part by 
the European Research Council (ERC) under the European Union's Horizon 
2020 research and innovation programme, grant agreement No 668679.





\bibliographystyle{elsarticle-num}
\bibliography{sjostrand}







\end{document}